# Evidence for the weakly coupled electron mechanism in an Anderson-Blount polar metal

N. J. Laurita[1,2], A. Ron[1,2], Jun-Yi Shan[1,2], D. Puggioni[3], N. Z. Koocher[3], K. Yamaura[4], Y. Shi[5], J. M. Rondinelli[3], & D. Hsieh[1,2]

[1]*Department of Physics, California Institute of Technology, Pasadena, California 91125, USA.*

[2]*Institute for Quantum Information and Matter, California Institute of Technology, Pasadena, California 91125, USA.*

[3]*Department of Materials Science and Engineering, Northwestern University, Illinois 60208-3108, USA.*

[4]*Research Center For Functional Materials, National Institute for Materials Science, 1-1 Namiki, Tsukuba, Ibaraka, 305-0044, Japan.*

[5]*Institute of Physics, Chinese Academy of Sciences, Beijing 100190, China.*

## Abstract

Over 50 years ago, Anderson and Blount proposed that ferroelectric-like structural phase transitions may occur in metals, despite the expected screening of the Coulomb interactions that often drive polar transitions. Recently, theoretical treatments have suggested that such transitions require the itinerant electrons be decoupled from the soft transverse optical phonons responsible for polar order. However, this decoupled electron mechanism (DEM) has yet to be experimentally observed. Here we utilize ultrafast spectroscopy to uncover evidence of the DEM in $LiOsO_3$, the first known band metal to undergo a thermally driven polar phase transition ($T_c$ =140 K). We demonstrate that intra-band photo-carriers relax by selectively coupling to only a subset of the phonon spectrum, leaving as much as 60% of the lattice heat capacity decoupled. This decoupled heat capacity is shown to be consistent with a previously undetected and partially displacive TO polar mode, indicating the DEM in $LiOsO_3$.

## Introduction

Ferroelectric transitions – in which a crystal spontaneously develops a switchable electric polarization – are typically driven by long-range Coulomb interactions and are therefore conventionally found in insulating dielectrics where such fields are unimpeded[1]. In metals, these interactions are immediately screened by the itinerant electrons, seemingly suggesting an incompatibility between metallicity and polarity[2]. However, recent theoretical treatments[2–4] have suggested an alternative route by which metals may achieve polar order, where polar instabilities are instead driven by short-range interactions originating from the local bonding environment of the cations in the unit cell. An experimental signature of these so-called geometric polar metals is naturally encoded in their electron-phonon interactions, as the viability of this mechanism is believed to hinge on a decoupling[5] of the itinerant electrons from the displacive transverse optical (TO) polar phonons which drive the transition. This underlying concept was first noted in Anderson and Blount's seminal 1965 proposal[6], but recently recast by Puggioni and Rondinelli[4] as a guiding operational principle in the design of polar metals. Despite its fundamentality, this decoupled electron mechanism (DEM) has never been experimentally verified due to the scarcity of metals which display intrinsic polar transitions. Thus, the strength of itinerant electron – polar phonon interactions and the mechanism by which metals may undergo polar transitions is currently unresolved.

An excellent testbed for uncovering the nature of itinerant electron – polar phonon interactions, and by extension the DEM, is LiOsO$_3$. This material shares an identical crystal structure to LiNbO$_3$-type ferroelectrics[7] and exhibits an analogous polar transition from the centrosymmetric $R\bar{3}c$ (Fig. 1a) to polar $R3c$ (Fig. 1b) space groups driven primarily by Li ion displacement along the trigonal [001] polar axis[8,9]. However, unlike LiNbO$_3$, LiOsO$_3$ is metallic in both the non-polar and polar phases[8]. Density functional theory calculations (Fig. 1c) suggest that this metallicity derives from the presence of O 2$p$ and possibly correlated[8–10] Os 5$d$ $t_{2g}$ orbitals at the Fermi level (see Supplementary Note 1). The fact that the polar transition is driven by Li ion displacement while the metallicity derives from O and Os orbitals is suggestive that the DEM may occur in LiOsO$_3$. However, the expected $A_{2u}$ TO soft mode[11] associated with the Li ion displacements was not detected by Raman spectroscopy[12], leaving both its coupling to the itinerant electrons and the displacive versus order-disorder character of the transition debated[9,10,13,14].

Ultrafast optical pump-probe experiments are capable of ascertaining how efficiently photo-generated carriers relax via various phonon decay channels[15,16] and are therefore well-suited to study how the electron phonon coupling strength varies across different phonon modes in LiOsO$_3$ (Fig. 1d,e). Here we utilize this technique to uncover evidence of the DEM in LiOsO$_3$. In our experiment, a pump photon energy of 1.56 eV was chosen so as to only generate photo-excitations within the metallic band, presumably via dipole allowed Os 5$d$ - O 2$p$ transitions (Fig. 1c), while still exceeding the maximum phonon energy of LiOsO$_3$[12] so as to avoid any restrictions on the photo-carrier -phonon scattering phase space. The phonon-mediated photo-carrier relaxation dynamics were then tracked via the pump induced fractional change in reflectivity ($\Delta R/R$), as measured by a time-delayed probe pulse of tunable energy, although the relaxation dynamics were not found to vary significantly within our accessible energy range (see Supplementary note 2).

**Results**

**Temperature dependence of the relaxation dynamics**

The measured temperature dependent reflectivity transients of LiOsO$_3$ are shown in Figure 2a. At all temperatures, $\Delta R/R$ displays an abrupt drop at time $t = 0$ followed by a recovery on the picosecond time scale to a negatively offset value. Both the magnitude of the drop and the recovery dynamics are clearly sensitive to $T_c$. To better highlight the temperature dependent photo-carrier relaxation dynamics, we subtract the offset from each transient and normalize their magnitudes (Figs. 2b,c). For $T \gg T_c$, only a single fast relaxation process with a time constant of $\tau_f \approx 0.1$ ps is observed, a typical timescale for electronic relaxation in metals[16]. However, as $T_c$ is approached, an additional slow relaxation process with time constant $\tau_s \approx$ 1-3 ps emerges and peaks in magnitude near the polar transition. As we demonstrate below, the emergence of this slower relaxation process is the result of a decoupled TO polar mode which displacively softens across $T_c$.

To capture the temperature dependence of these two relaxation processes, we fit the reflectivity transients to a phenomenological bi-exponential model given by $\Delta R/R = A_f \exp(-t/\tau_f) + A_s \exp(-t/\tau_s) + C$, where $A_f$ and $A_s$ are the amplitudes of the fast and slow relaxation components respectively and the constant offset $C$ accounts for slow ($\approx$ 20 ns) heat diffusion out of the probed region of the sample (see Supplementary Notes 4 and 5). We begin by highlighting the temperature dependence of $\tau_f$ (Fig. 2d) which is found to closely track the structural order parameter as captured by the anisotropic thermal parameter $\beta_{33}$[8]. While this resemblance would appear to suggest that the relaxation dynamics are strongly tied to the polar transition, the order-parameter-like increase of $\tau_f$ is relatively modest ($\approx$ 25%). This is far weaker, for instance, than the $\approx$ 500% divergence in photo-carrier lifetimes exhibited by the parent compounds of the pnictide superconductors at their structural transitions[17,18], and therefore not necessarily inconsistent with the DEM. Indeed, weak coupling between the electronic structure and the polar transition is further supported by the temperature dependence of $A_\text{f}$ (Fig. 2d inset), which is a measure of the change in the joint density of states at the probe wavelength and therefore exceptionally sensitive to subtle variations in the electronic structure. Despite this sensitivity, $A_\text{f}$ is found to exhibit weak temperature dependence across $T_c$, exemplifying the insensitivity of the low energy band structure to the polar transition, as expected from first principles calculations[9,11] and consistent with the DEM[4,6].

In contrast to the fast relaxation component, the slow relaxation component displays stronger temperature dependence, as exemplified by both $A_\text{s}$ and $\tau_\text{s}$ which exhibit a cusp-like behavior across $T_\text{c}$ (Fig. 2e). However, as will be demonstrated below, this slow relaxation component does not result from electron-phonon interactions. This is in accordance with the general expectations of intra-band photo-carrier relaxation in metals, as the existence of multiple phonon decay channels in a hot metal does not result in separate relaxation processes, but instead, a single relaxation process with a relaxation rate given by a Matthiessen-type rule as $1/\tau_\text{tot} = 1/\tau_1 + 1/\tau_2 + \ldots$ etc. Instead, the emergence of a second relaxation component in metals is generally indicative of coupling to additional bosonic modes (e.g. magnons[19]), a gap induced inter-band relaxation bottleneck[20], or a preferred electron-phonon coupling in which only a subset of the phonon spectrum mediates photo-carrier relaxation[21].

**Microscopic origin of the relaxation dynamics**

To rule out a gap induced inter-band relaxation bottleneck, we performed measurements as a function of pump fluence, which are capable of distinguishing between single particle and bi-molecular relaxation processes[22,23]. Figure 3 displays the fluence dependence of the parameters of the bi-exponential model (see Supplementary Note 6). The amplitudes $A_\text{f}$ and $A_\text{s}$ (Fig. 3a,b) both exhibit the linear fluence dependence expected of the linear response regime. However, both $\tau_f$ and $\tau_s$ (Fig. 3c,d) are fluence independent within our error bars, which is inconsistent with the linear fluence dependence expected from bi-molecular recombination dynamics across a gap[23], and thus rules out a gap induced interband relaxation bottleneck as the origin of the slow relaxation process. With no other collective excitations aside from phonons known in LiOsO$_3$[8], we conclude that a preferred electron-phonon coupling exists, in which photo-carriers selectively couple to only a subset of the overall phonon spectrum - referred to here as the strongly coupled phonons (SCPs) - before thermalization with the rest of the phonon modes - the weakly coupled phonons (WCPs) - occurs.

To identify which phonon modes strongly and weakly couple to excited photo-carriers, we appeal to a three-temperature thermalization model (TTM), which captures the selective electron-phonon coupling dynamics by treating the electrons, SCPs, and WCPs as three coupled thermal baths (Fig. 3e)[21]. In the TTM, excited photo-carriers, which have rapidly thermalized to a high electronic temperature (see Supplementary Note 7), relax by thermalizing with the SCPs via electron-phonon coupling $g_{\text{ep}}$, resulting in a single relaxation process with time constant $\tau_{\text{f}}$. These SCPs then thermalize with the WCPs via anharmonic phonon coupling $g_{\text{pp}}$, prompting an additional relaxation process with time constant $\tau_{\text{s}}$. The temperature dependent heat capacities of the two lattice thermal baths are constructed by partitioning the reported[8] total lattice heat capacity $C_{\text{p}}$ by the parameter $\alpha < 1$, such that the SCPs carry heat capacity $C_{\text{s}} = \alpha C_{\text{p}}$ while the WCPs carry heat capacity $C_{\text{w}} = (1 - \alpha)C_{\text{p}}$. By solving the TTM under the experimentally defined initial conditions, we obtain the transient electronic ($T_{\text{e}}$), SCP ($T_{\text{s}}$), and WCP ($T_{\text{w}}$) temperatures (Fig. 4a), which are then combined via a conventional[16,21] weighted sum to form model reflectivity transients as

$$\frac{\Delta R}{R} = aT_{\text{e}} + b[\alpha T_{\text{s}} + (1 - \alpha)T_{\text{w}}] \qquad (1)$$

where *a* and *b* are determined by the initial and final values of the experimental data. These model reflectivity transients are then fit to the experimental data (Fig. 4b), allowing for unique extraction of $g_{ep}$, $g_{pp}$, and $\alpha$ as fitting parameters (see Supplementary Note 8).

**Identification of the strongly and weakly coupled phonons**

With the thermalization model applied, we may now determine which phonon modes constitute the SCPs and WCPs. We begin by identifying the SCPs via a comparison of the extracted electron-phonon coupling function $g_{ep}$ to the phonon linewidths $\Gamma_{ph}$ of LiOsO$_3$. In first principles electron-phonon coupling theory[5], the strength of the coupling between the electronic structure and a particular phonon mode is naturally encoded in the phonon's lifetime. In the zero momentum limit, this manifests as a $g_{ep} \propto \sqrt{\hbar\Gamma_{ph}}$ scaling relation, thereby providing a route to identifying which modes primarily mediate photo-carrier relaxation. Among the phonons reported by Raman spectroscopy[12], the $^1E_g$ and $^2E_g$ modes exhibit a temperature dependent linewidth that appear to obey this scaling relation (Fig. 4c) (see Supplementary Note 9). This suggests that these modes, and possible others whose linewidths have not yet been reported, constitute the SCPs. We can ascertain the coupling strength to these modes by converting $g_{ep}$ into the dimensionless form $\lambda$ via the relation $g_{ep} = (6\hbar\gamma/\pi k_B)\lambda\langle\omega^2\rangle$, where $\gamma$ is the Sommerfeld coefficient and $\langle\omega^2\rangle$ is the second moment of the $^1E_g$ and $^2E_g$ phonon frequencies[15]. Through this analysis, we find a dimensionless coupling of $\lambda = 0.09$ at the polar transition of LiOsO$_3$, a value comparable to that of more conventional metals[24]. It should be noted that the $^1E_g$ and $^2E_g$ modes are primarily associated with distortions of the OsO$_6$ octahedra[12] (Fig. 4c inset) and are thus not associated with Li ion motion along the polar axis, consistent with the DEM.

To identify which modes constitute the WCPs, we examine the temperature dependence of $1 - \alpha$ (Fig. 4d), which represents the heat capacity of these weakly coupled phonons. Before proceeding, it should be noted that the presence of a selective electron-phonon coupling in LiOsO$_3$, i.e. an $\alpha < 1$, is in itself peculiar. Previously, such selective coupling has only been observed in materials such as graphite[21], iron pnictides[22], and cuprates[25] and has been attributed to their reduced effective dimensionality[26]. Essentially, their layered structures naturally give rise to a preferred coupling to in-plane rather than inter-plane phonon modes, resulting in a spatially anisotropic electron-phonon coupling. However, LiOsO$_3$ does not possess a layered structure and there has thus far been no evidence that it behaves as an effective 2D system[9], suggesting a distinct explanation for the observed selective electron-phonon coupling. Instead, the selective coupling is naturally explained by the DEM, in which a weakly coupled polar mode first softens and then hardens across a partially displacive-like polar transition.

To demonstrate this, we show that the temperature dependence of $1-\alpha$ is accounted for by the heat capacity of a displacive polar mode. We restrict this discussion to temperatures $T < 250$ K, below which the non-polar optical modes are expected to be frozen out[12]. Therefore, in this regime the polar mode is the only thermally populated optical mode and we can approximate $1-\alpha \approx C_{A_{2u}}(T)/C_p(T)$, where $C_{A_{2u}}(T)$ is the heat capacity of the $A_{2u}$ polar mode. We can then model $1-\alpha$ by treating the polar mode as an Einstein phonon whose temperature dependent frequency $f_{A_{2u}}$ is then the only free parameter. Despite the simplicity of this model, we find that it not only completely reproduces the temperature dependence of $1-\alpha$ (black line in Fig. 4d), but also allows for the extraction of the temperature dependent heat capacity and frequency of the polar mode (Fig. 4d inset), which clearly shows the cusp-like behavior across $T_c$ emblematic of displacive polar phonons (see Supplementary Note 10). The validity of this interpretation is further supported by the fact that the functional dependence of the extracted polar mode frequency below $T_c$ is well captured by the Cochran relation[27] $\hbar\omega \propto \sqrt{1-T/T_c}$ expected of polar soft modes. Furthermore, the extracted polar mode frequency of $f_{A_{2u}} \approx 7$ THz at our lowest measured temperature is in excellent agreement with zero temperature calculations, which predict a polar mode frequency of $f_{A_{2u}} \approx 7.2$ THz (see Supplementary Note 11). This analysis not only demonstrates the partially displacive character of the transition, as opposed to the strictly order-disorder mechanism proposed due to the absence of a Li soft mode in the Raman spectra[12], but also shows that the photo-carriers couple extremely weakly to the polar mode (i.e. the polar mode belongs to the set of WCPs), thus indicating the DEM in LiOsO$_3$.

## Discussion

Having presented evidence for the DEM to occur in LiOsO$_3$, we now discuss the ramifications for polar materials in general. The DEM is contingent on a particular form of the electron-phonon coupling, in which the itinerant electrons are prevented from coupling to TO phonons by the polarization factor **q** · **e**$_\mathbf{q}$ in the electron-phonon coupling matrix elements, where **q** and **e**$_\mathbf{q}$ are the phonon wave vector and polarization respectively[5]. Our results suggest this form of electron-phonon coupling to be an excellent approximation in LiOsO$_3$, and may be largely applicable to polar metals. However, while the itinerant electrons appear to be nearly decoupled from the polar transition, we cannot rule out a small but finite coupling that perhaps contributes to the modest 25% increase in $\tau_f$ in the polar phase (Fig. 2d). One way that such finite coupling may arise is from the longitudinal optical (LO) / TO degeneracy due to the screened Coulomb interactions in polar metals, which was not accounted for in Anderson and Blount's original proposal[6] but is captured by Puggioni and Rondinelli's weak-coupling operational principles[4]. At finite carrier densities the LO/TO modes mix as $\mathbf{k} \to 0$ and thus a unique differentiation between the LO/TO modes participating in the loss of inversion symmetry is no longer possible (see Supplementary Note 11). Taken together, our experimental results support a picture in which polar transitions in metals are driven by short-range interactions[2,13] related to the bonding environment of the cations within the unit cell, which endure the metallicity by virtue of being decoupled from the electronic structure at the Fermi level.

## Methods

**Time-Resolved reflectivity measurements**. Time-resolved reflectivity experiments were performed using a pump pulse with center wavelength 795 nm (1.56 eV) and duration ≈100 fs produced by a regeneratively amplified Ti:sapphire laser system operating at a 100 kHz repetition rate. The probe pulse was produced by an optical parametric amplifier operating at the same repetition rate. By referencing a lock-in amplifier to the frequency that the pump beam is mechanically chopped (10 kHz), fractional changes in the reflectivity $\Delta R/R$ as small as 10$^{-5}$ can be resolved. Temperature dependent measurements were performed with a pump pulse of fluence $F = 0.5$ mJ/cm$^2$ while the probe pulse had center wavelength 1350 nm (0.92 eV) and fluence $F = 10$ µJ/cm$^2$. Fluence dependent measurements were performed at $T = 80$ K by varying the pump pulse fluence while the probe pulse was maintained at center wavelength 1500 nm (0.83 eV) and fluence $F = 20$ µJ/cm$^2$. All pulses were focused at near normal incidence on the [42$\bar{1}$] face of a single crystal sample of approximate dimensions 0.25 mm x 0.5 mm x 0.25 mm grown by a solid state reaction under pressure (see Ref. 8 for details regarding sample preparation and characterization).

**Three-temperature model of the relaxation dynamics.** The three-temperature model assumes that excited photo-carriers thermalize with a set of strongly coupled phonons before thermalization with the rest of the lattice occurs. In the model, the total lattice heat capacity $C_p$ is partitioned into two separate phononic thermal baths, such that the strongly and weakly coupled phonons carry heat capacities $C_s = \alpha C_p$ and $C_w = (1-\alpha) C_p$ respectively, where the parameter $\alpha < 1$. In this fashion, $\alpha$ describes the portion of the total lattice heat capacity which participates in photo-carrier - lattice thermalization, and may thus be used to determine which modes couple strongly to the excited photo-carriers.

In the model, excited photo-carriers are assumed to immediately thermalize to a Fermi-Dirac distribution at a high electronic temperature given by[21]

$$T_{e,i} = \frac{1}{\delta_s} \int_0^{\delta_s} \left[ \sqrt{T_i^2 + \frac{2(1-R)F}{\delta_s \gamma} \exp\left(-\frac{z}{\delta_s}\right)} \right] dz \qquad (2)$$

where $T_i$ is the temperature before pump excitation, $\gamma$ is the Sommerfeld coefficient, $R$ is the reflectivity at the pump wavelength, $F$ is the pump fluence, $z$ is the depth into the sample, and integration is performed over one penetration depth $\delta_s$ at the pump wavelength. We estimate photo-carrier thermalization occurs within a few fs after pump excitation[28] (see Supplementary Note 7). Heat exchange between the electronic and two lattice thermal baths is then governed by the equations

$$2C_e \frac{\partial T_e}{\partial t} = -g_{ep}(T_e - T_s) + I(t,z) + \nabla \cdot [\kappa_e \nabla T_e] \qquad (3)$$

$$C_s \frac{\partial T_s}{\partial t} = g_{ep}(T_e - T_s) - g_{pp}(T_s - T_w) \qquad (4)$$

$$C_w \frac{\partial T_w}{\partial t} = g_{pp}(T_s - T_w) \qquad (5)$$

where $C_e$ is the electronic heat capacity, $I(z,t)$ is the laser source term, $\kappa_e$ is the thermal conductivity, $g_{ep}$ and $g_{pp}$ are the electron-phonon and phonon-phonon coupling functions, and $T_s$ and $T_w$ are the temperatures of the strongly and weakly coupled phonons respectively.

The three temperature model equations are then solved to obtain the time dependent electronic and lattice temperatures. Model reflectivity transients are then constructed by convolving a normalized Gaussian with a conventional[16,21] weighted sum of the electronic and lattice temperatures as

$$\frac{\Delta R}{R} = aT_e + b[\alpha T_s + (1-\alpha)T_w] \tag{6}$$

where $a$ and $b$ are determined by initial and final values of the experimental reflectivity transients. The model reflectivity transients are then fit to the experimental data using a least squares regression algorithm with $g_{ep}$, $g_{pp}$, and $\alpha$ as relaxed fitting coefficients (see Supplementary Note 8).

**Data availability**

The datasets generated are/or analysed during the current study are available from the corresponding author on reasonable request.

**Additional information**

Correspondence and requests for materials should be addressed to D.H. (dhsieh@caltech.edu).

**Acknowledgements**

This work was supported by the U.S. Department of Energy under Grant No. DE SC0010533. D. H. also acknowledges funding from the David and Lucile Packard Foundation and support for instrumentation from the Institute for Quantum Information and Matter, an NSF Physics Frontiers Center (PHY-1125565) with support of the Gordon and Betty Moore Foundation through Grant No. GBMF1250. N. J. L. acknowledges support from the Institute for Quantum Information and Matter Postdoctoral Fellowship. N. Z. K. was supported by U.S. DOE-BES grant no. DE-SC0012375. D. P. and J. M. R. were supported by ARO (Award No. W911NF-15-1-0017). Y. G. S was supported by the National Key Research and Development Program of China (No. 2017YFA0302901, 2016YFA0300604). We thank Rohit Prasankumar for helpful conversations, Qingming Zhang for sharing unpublished Raman data, and Stefano Lupi for sharing unpublished reflectivity results.

**Author contributions**

D.H. and N.J.L conceived the experiment. N.J.L. and A.R. performed the time-resolved reflectivity measurements. J.S. determined the crystal alignment. D.P., N.Z.K., and J.M.R. performed the DFT calculations, discussed the data and soft mode theories, and commented on the manuscript. Y.S. and K. Y. prepared and characterized the sample. N.J.L and D.H. analysed the data. N.J.L. and D.H wrote the manuscript.

**Competing Interests**

The authors declare no competing interests.

**Figure 1**

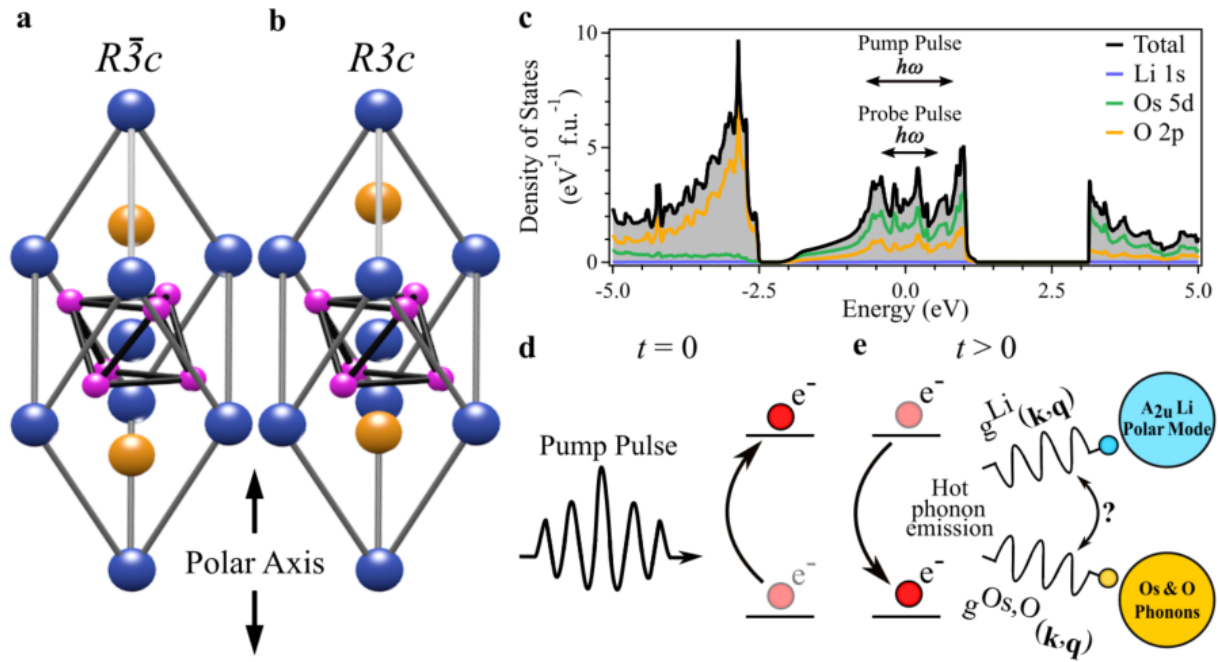

**Scheme for probing electron-phonon interactions of LiOsO$_3$** Crystal structure of LiOsO$_3$ in the **a,** non-polar $R\bar{3}c$ and **b,** polar $R3c$ phases with blue, orange, and pink spheres representing Os, Li, and O atoms respectively. These phases are distinguished by the displacement of the Li ions along the polar axis (black arrows). **c,** Density functional theory calculation of the orbital resolved electronic density of states of the $R3c$ structure of LiOsO$_3$ (see Supplementary Note 1). Our pump (1.56 eV) and probe (0.92 eV) pulse energies permit excitation and monitoring of only intra-band photo-excitations, as shown by the black arrows in the inset. **d,e** A cartoon of our experimental scheme for probing itinerant electron – polar phonon interactions in LiOsO$_3$. **d,** An intense pump pulse generates electronic excitations within the metallic band of LiOsO$_3$ at an initial time $t = 0$. **e,** These excitations relax at later times $t > 0$ by coupling to the lattice, naturally embedding the strength of the electron phonon coupling $g(\mathbf{k},\mathbf{q})$ in their relaxation rate. By comparing the temperature dependent relaxation rate to that expected of polar phonons, we may identify which phonon modes, either the $A_{2u}$ polar mode (blue sphere) or O and Os modes (orange sphere), primarily mediate photo-carrier relaxation and therefore are most strongly coupled to the itinerant electrons of LiOsO$_3$.

**Figure 2**

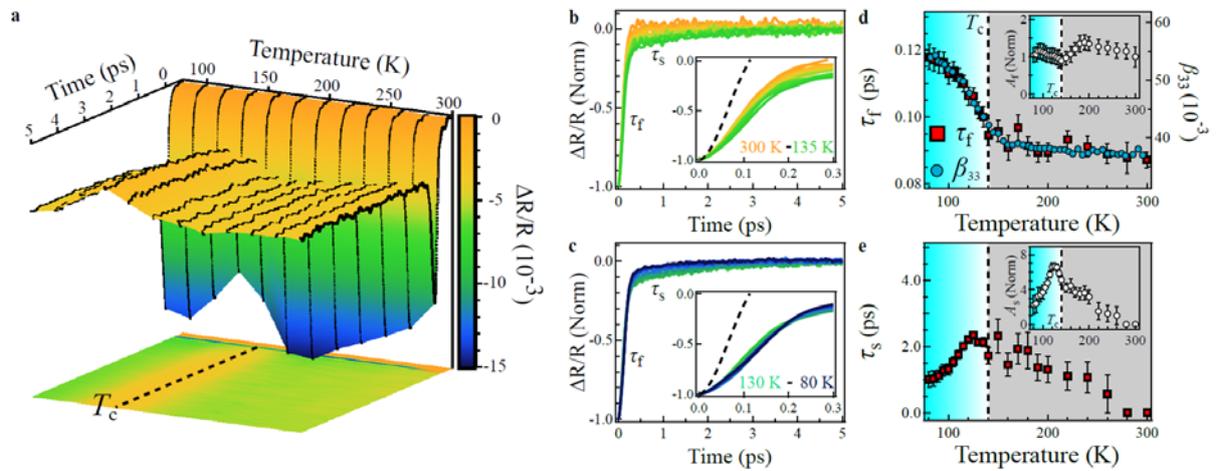

**Temperature dependent relaxation dynamics of LiOsO$_3$ a,** Three-dimensional surface plot of the transient reflectivity $\Delta R/R$ of LiOsO$_3$ as a function of temperature and time delay. Black lines are raw traces at select temperatures. An image plot of this data is projected at the bottom where clear signatures of the polar transition are observed at $T_c$ = 140 K. **b,c** Normalized reflectivity transients after background subtraction showing two distinct relaxation timescales, a faster relaxation with time constant $\tau_f$ and a slower relaxation with time constant $\tau_s$, in the **b,** non-polar and **c,** polar phases. Insets: Magnified view of the fast decay process compared to the Gaussian instrument resolution of our experiment, shown as a black dashed line, demonstrating our experiment is not resolution limited (see Supplementary Note 3). **d,e** Results of modeling the reflectivity transients with a bi-exponential function. **d,** Temperature dependence of $\tau_f$ plotted with the anisotropic displacement parameter $\beta_{33}$, a measure of the structural order parameter[8]. Inset: Temperature dependence of the amplitude of the fast relaxation $A_f$ normalized by its 300 K value. **e,** Temperature dependence of $\tau_s$. Inset: Temperature dependence of the slow relaxation $A_s$ normalized by its 260 K value, the temperature at which it is first discernable. Error bars in **d,e** derive from the $\chi^2$ of the bi-exponential fits.

**Figure 3**

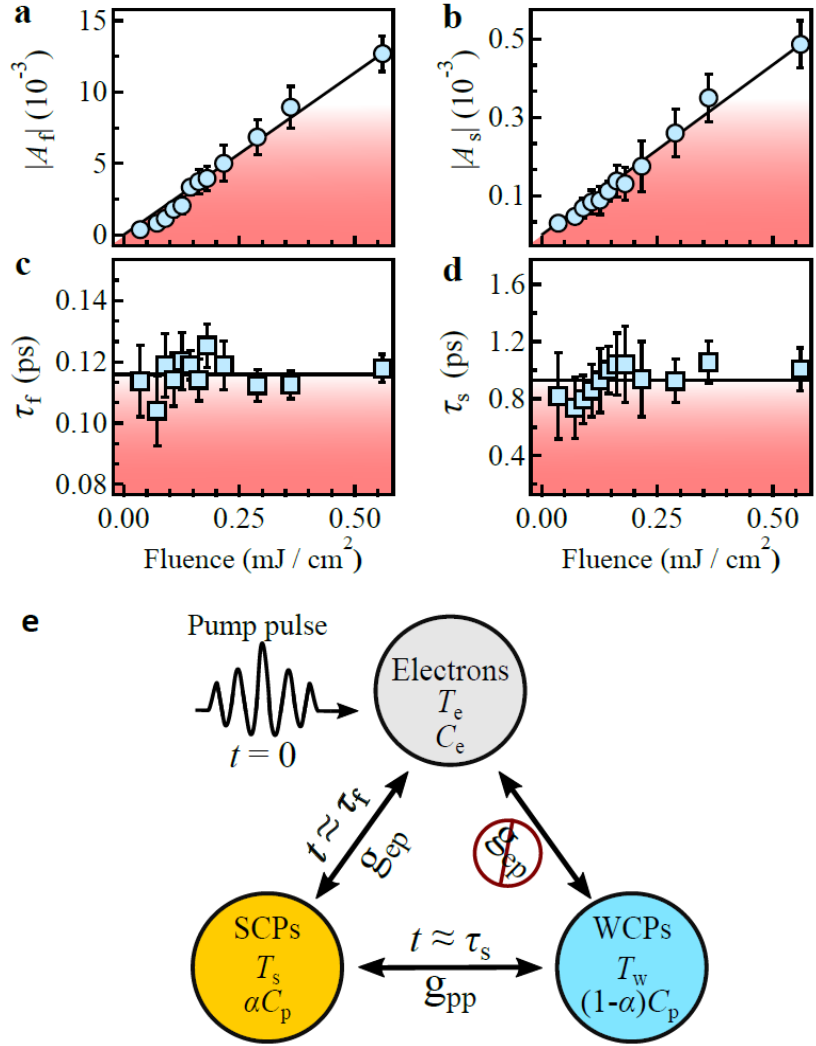

**Microscopic origin of the two relaxations in LiOsO₃ a-d,** Fluence dependence of the transient reflectivity of LiOsO₃ at $T$ = 80 K as captured by the parameters of the bi-exponential model. Both the amplitudes of the **a,** fast and **b,** slow relaxation processes display a linear dependence that is characteristic of the linear response regime. However, the decay constants of the **a,** fast and **b,** slow relaxation processes display a lack of fluence dependence that is indicative of intra-band photo-carrier relaxation and therefore suggests a selective electron-phonon coupling in LiOsO₃. Black lines in **a-d** are linear fits of the data while error bars derive from the $\chi^2$ of the bi-exponential fits. **e,** Schematic of microscopic origin of the relaxation processes of LiOsO₃. Electrons, initially excited to a high effective temperature $T_e$ by the pump pulse, relax on the timescale $\tau_f$ by thermalizing with only a set of strongly coupled phonons (SCPs) at temperature $T_s$ via electron-phonon coupling $g_{ep}$. The SCPs then relax on a time scale $\tau_s$ by thermalizing with the rest of the lattice modes, referred to as the weakly coupled phonons (WCPs) at temperature $T_w$, via phonon-phonon coupling $g_{pp}$. The relative heat capacities of the two phononic thermal baths are found by partitioning the total lattice heat capacity $C_p$ by the parameter $\alpha$.

**Figure 4**

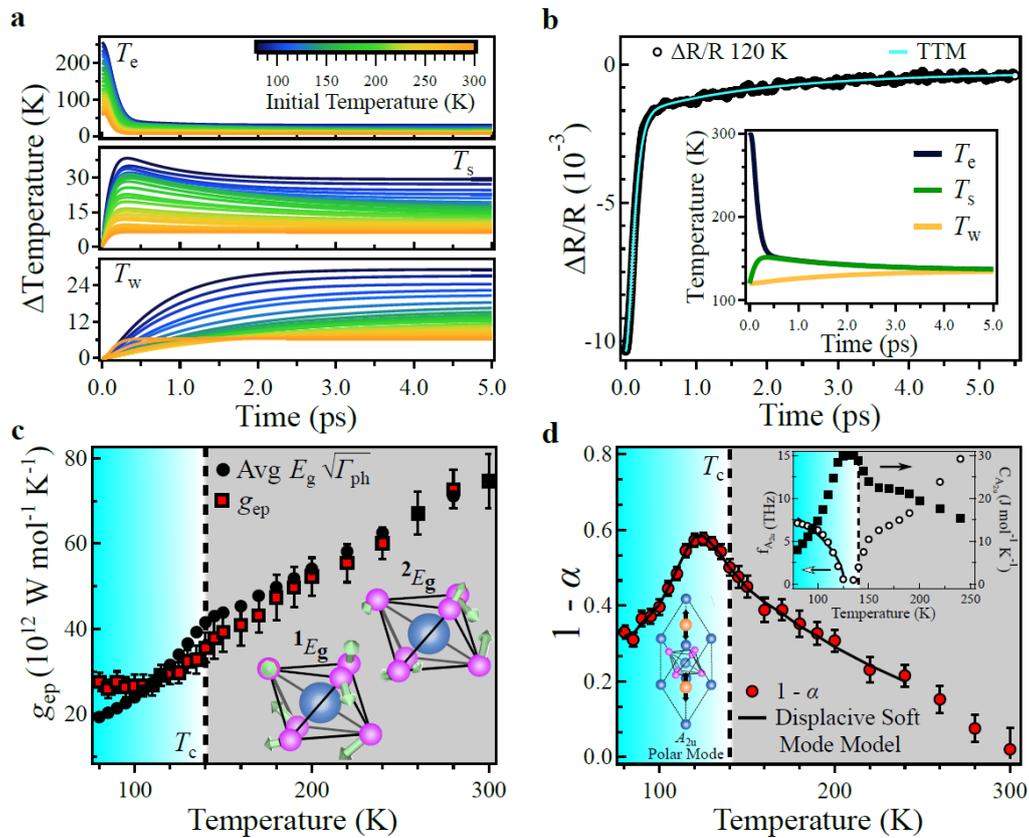

**Identification of the strongly and weakly coupled phonons a,** Transient changes in the electronic ($T_e$), strongly coupled phonon ($T_s$), and weakly coupled phonon ($T_w$) temperatures as extracted from a three-temperature thermalization model (TTM) of the relaxation dynamics of LiOsO$_3$. Color scale represents different initial experiment temperatures before the pump pulse arrives. **b,** Representative fit of an experimental reflectivity transient at an initial temperature of $T$ = 120 K (Black circles) produced by the TTM (teal line). Inset: Transient temperatures $T_e$, $T_s$, and $T_w$ from which the model reflectivity transient is generated. **c,** Comparison of the extracted electron-phonon coupling function $g_{ep}$ with an average of the square roots of the $^1E_g$ and $^2E_g$ phonon linewidths $\Gamma_{ph}$ as reported by Raman spectroscopy[12], suggesting these modes are strongly coupled phonons. Inset: Real space distortions of these modes with blue and pink spheres representing Os and O atoms respectively. **d,** Temperature dependence of the heat capacity of the weakly coupled phonons represented by 1 - $\alpha$. The dashed vertical line denotes the reported value of $T_c$. The solid black line is a fit to a model where this heat capacity is attributed to a displacive $A_{2u}$ polar mode (lower inset) that softens and hardens across the polar transition. Upper inset: Resonant frequency $f_{A_{2u}}$ (open circles) and heat capacity $C_{A_{2u}}$ (closed squares) of the polar mode extracted from the displacive soft mode model. The solid black line is a fit of the resonant frequency in the polar phase to the Cochran relation[27] $\hbar\omega \propto \sqrt{1 - T/T_c}$ expected of polar soft modes, in which $T_c$ was shifted to coincide with the peak of 1 – $\alpha$ at $T_c$ = 125 K. All error bars derive from the $\chi^2$ of the three-temperature model fits.

Supplementary Information For

# Evidence for the weakly coupled electron mechanism in an Anderson-Blount polar metal

N. J. Laurita et al.

## Supplementary Note 1: Details of the density functional theory calculations

We performed electronic structure calculations for the $R3c$ ground state structure of $LiOsO_3$ based on density functional theory (DFT) within the revised Perdew–Burke–Ernzerhof exchange-correlation functional revised for solids[1] as implemented in the Vienna Ab initio Simulation Package[2] with the projector-augmented wave method[3] to treat the core and valence electrons using the following electronic configurations $1s^2\ 2s^1$ (Li), $6s^1\ 5d^7$ (Os), and $2s^2 2p^4$ (O), and a 550 eV plane wave cutoff. An $11 \times 11 \times 11$ Monkhorst-Pack $k$-point mesh[4] and Gaussian smearing of 20 meV width was used for the Brillouin zone sampling and integrations. For structure optimization we relaxed the lattice constants and the atomic positions to have forces less than 0.1 meV Å$^{-1}$. The crystallographic parameters of the relaxed $R3c$ structure are given in Supplementary Table I while the crystal structure is shown in Figure 1a of the main text.

| Atom | x | y | z | Wyckoff position | Site symmetry |
|------|---------|---------|---------|------------------|---------------|
| Li   | 0.00000 | 0.00000 | 0.21443 | 6a               | $D_3$         |
| Os   | 0.00000 | 0.00000 | 0.00013 | 6a               | $S_6$         |
| O    | 0.00339 | 0.63367 | 0.25255 | 18b              | $C_2$         |

Supplementary Table I: **Crystallographic parameters of the $R3c$ phase** Wyckoff positions are given in units of the lattice constants.

## Supplementary Note 2: Demonstration that the photo-carrier dynamics are not strongly energy dependent

To ensure that the photo-carrier dynamics of $LiOsO_3$ are not strongly energy dependent, we performed time-resolved reflectivity experiments as a function of probe pulse energy. The experiment was conducted at $T = 80$ K with the pump pulse energy maintained at 1.56 eV and fluence $F = 0.5$ mJ/cm$^2$ while the probe pulse energy was varied with the fluence maintained at $F = 20$ μJ/cm$^2$. Supplementary Figure 1 displays the normalized measured reflectivity transients as a function of probe energy with fits to the bi-exponential model (dashed lines). One can see that the relaxation dynamics do not vary significantly throughout the low energy manifold of states of $LiOsO_3$.

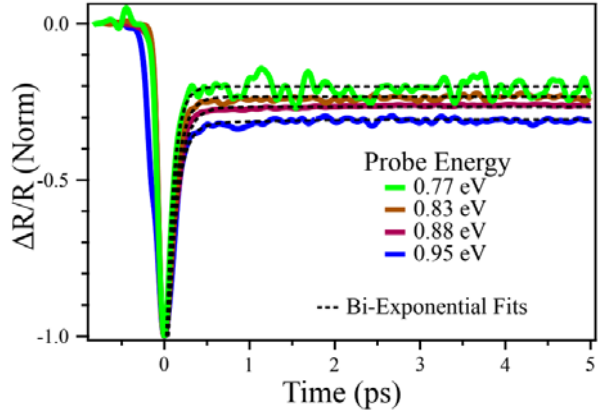

Supplementary Figure 1: **Probe energy dependent relaxation dynamics** Normalized reflectivity transients taken at $T = 80$ K as a function of probe pulse energy. Black dashed lines are fits to the bi-exponential model.

## Supplementary Note 3: Demonstration that the experiment is not resolution limited

The Gaussian instrument resolution of the experiment can be determined by fitting the acquired normalized reflectivity transients with a Gaussian function for times $t < 0$ and projecting this fit to times $t > 0$. This fit represents the fastest response that the experiment can be expected to reliably measure. Supplementary Figures 2**a-f** display the measured reflectivity transients, normalized by their negative peak values, at several representative temperatures. Also shown is the Gaussian instrument resolution as a black dashed line. One can see that the instrument resolution is faster than the relaxation of the measured reflectivity transients at all temperatures, demonstrating that our measurements are not resolution limited.

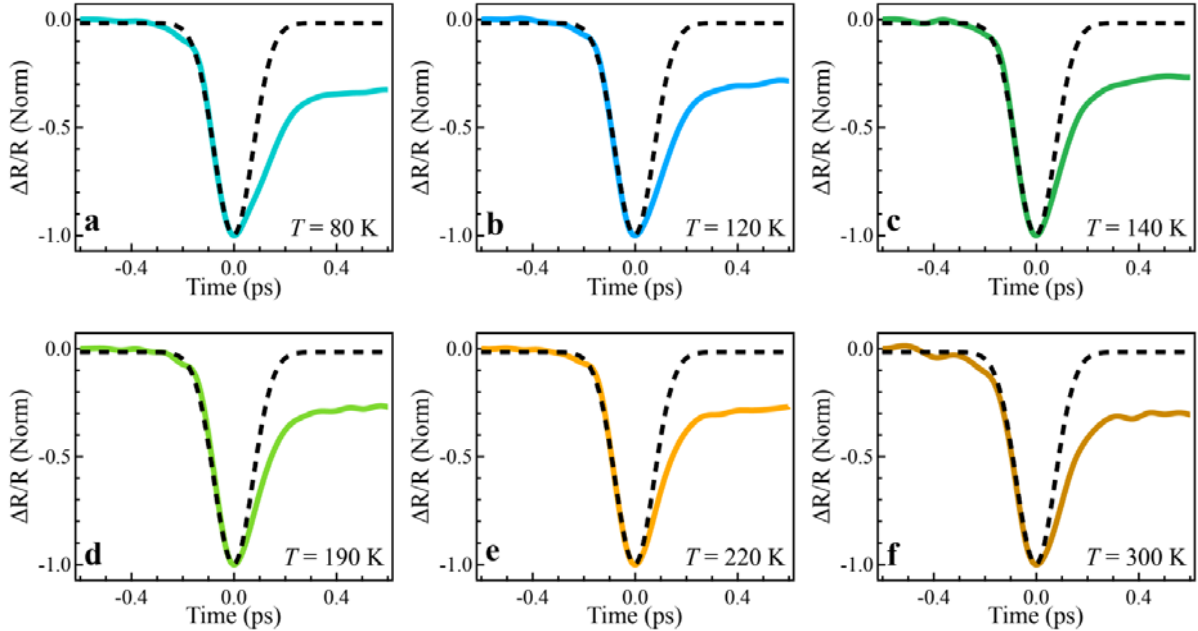

Supplementary Figure 2: **Demonstration that the experiment is not resolution limited a-f,** Normalized reflectivity transients at several representative temperatures with the Gaussian instrument resolution (black dashed line) superimposed on the data.

## Supplementary Note 4: Details of the bi-exponential fits

A qualitative understanding of the temperature dependence of the measured reflectivity transients were obtained by fitting the data to a bi-exponential model given by,

$$\frac{\Delta R}{R} = A_\text{f} \exp\left(-t/\tau_\text{f}\right) + A_\text{s} \exp\left(-t/\tau_\text{s}\right) + C \quad (1)$$

where $A_\text{f}$, $A_\text{s}$, $\tau_\text{f}$, and $\tau_\text{s}$ are the amplitudes and decay constants of the fast and slow relaxation components identified in $LiOsO_3$ respectively and $C$ is a constant which captures heat diffusion out of the probed region of the sample. Supplementary Figure 3 displays fits of the reflectivity transients to the bi-exponential model at two representative temperatures. Experimental data is shown as black circles while bi-exponential fits of the data are shown as teal dashed lines. Also shown are each component of the fit with the corresponding fitting coefficients listed in the bottom right. Supplementary Figure 3a displays the data and fit at $T = 300$ K, where we find that a single exponential is sufficient to describe the data. However, as the temperature is reduced, the slow relaxation component grows in amplitude and must be accounted for in the fit. Supplementary Figure 3b shows the data and fit at $T = 120$ K, a temperature near where the slow relaxation is maximum.

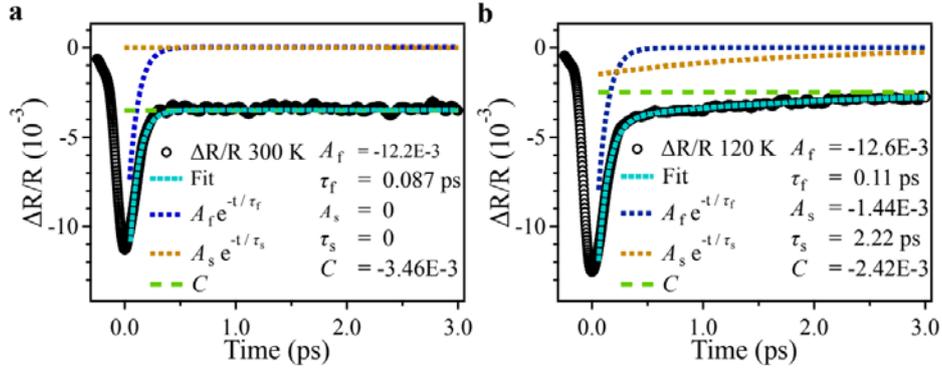

Supplementary Figure 3: **Sample fits to the bi-exponential model a,b** Reflectivity transients (black circles) at two representative temperatures, **a**, $T = 300$ K and **b**, $T = 120$ K, with the corresponding fits to the bi-exponential model (teal dashed lines). Individual components of the fits are shown as dashed lines while their corresponding fitting coefficients listed in the bottom right.

## Supplementary Note 5: Calculated heat diffusion timescale

The background term $C$ in the bi-exponential model (Supplementary Equation 1) describes heat diffusion into the bulk of the material to depths larger than the penetration depth at the probe wavelength. An estimate for this timescale is given by,

$$t_\text{c} = \frac{\delta_\text{s}^2 d C_\text{p}}{M \kappa_\text{e}} \quad (2)$$

where $\delta_\text{s}$ is penetration depth at the probe wavelength, $d$ is the density, $C_\text{p}$ is the lattice heat capacity, $M = 245.17$ g/mol is the molar mass, and $\kappa_\text{e}$ is the thermal conductivity[5].

Aside from the lattice heat capacity[6] these quantities have yet to be reported in $LiOsO_3$. However, estimates can be made using values from comparable materials or relations with already reported quantities. The density is estimated to be $d = 4.65$ g/cm$^3$ from isostructural $LiNbO_3$. An estimate of the thermal conductivity is provided by the Wiedemann-Franz law, which relates $\kappa_\text{e}$ to the resistivity as $\kappa_\text{e} = LT_\text{e}/\rho$, where $L = 2.44 \times 10^{-8}$ W $\Omega$ K$^{-2}$ is the Lorentz number. From the reported resistivity[6], we estimate $\kappa_\text{e} = 0.4$ W m$^{-1}$ K$^{-1}$ at $T = 100$ K. The resistivity can also be used to calculate the penetration depth at the probe wavelength (See Supplementary Equation 6 below), which we estimate to be $\delta_\text{s} \approx 100$ nm at $T = 100$ K and $\lambda = 1350$ nm.

Inserting these values into Supplementary Equation 2 results in an expected heat diffusion timescale of $t_\text{c} = 24$ ns. As this timescale is well in excess of our measurement time of 10 ps, $C$ is well approximated by a time-independent constant.

## Supplementary Note 6: Raw fluence dependent data

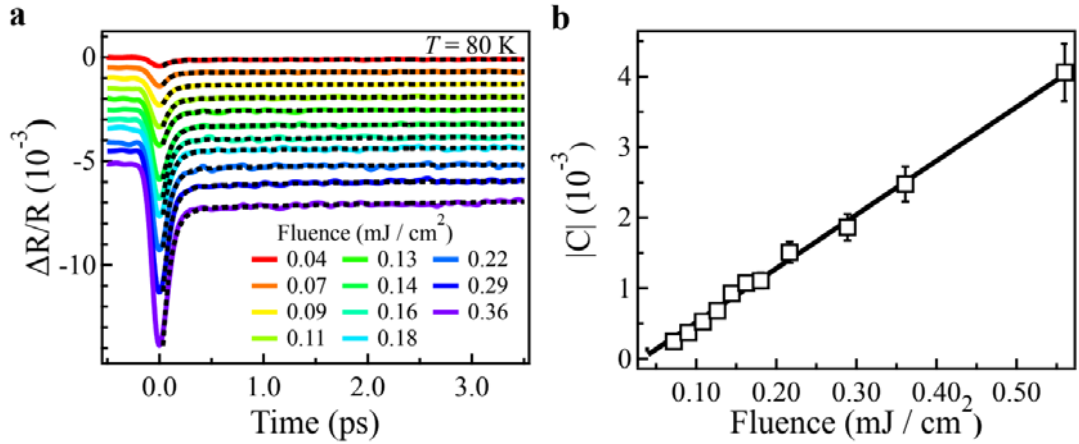

Supplementary Figure 4: **Fluence dependent relaxation dynamics a**, Reflectivity transients of LiOsO$_3$ as a function of pump fluence measured at $T = 80$ K with the pump pulse energy maintained at 1.56 eV while the probe pulse energy and fluence were maintained at 0.83 eV and $F = 20$ $\mu$J/cm$^2$ respectively. Curves have been offset vertically for clarity. Black dashed lines are fits to the bi-exponential model. **b**, Fluence dependence of the constant $C$ in the bi-exponential model. Error bars derive from the $\chi^2$ of the bi-exponential fits.

## Supplementary Note 7: Verification of rapid photocarrier thermalization

Thermalization models of transient reflectivity experiments require that the excited photocarriers initially thermalize to Fermi-Dirac distribution before thermalization with the lattice begins. To estimate the timescale for electron-electron thermalization in LiOsO$_3$, we follow the calculation outlined in Ref. 7.

In the context of Fermi liquid theory, the timescale needed for a Fermi gas to thermalize to a temperature $T$ can be approximated by,

$$t_{\text{th}} \approx \frac{1}{2K(k_B T)^2} \quad (3)$$

where $K$ is the electron-electron scattering rate derived from the angular averaged electron-electron scattering probability. An estimate for $K$ via the random phase approximation is given by,

$$K = \frac{\pi^2 \sqrt{3}}{128} \frac{\omega_p}{E_F^2} \quad (4)$$

where $\omega_p$ is the plasma frequency and $E_F$ is the Fermi energy.

The plasma frequency of LiOsO$_3$ was measured to be $\omega_p \approx 90$ THz in the reflectivity experiments of Lo Vecchio *et al.*[8,9]. An estimate for the charge density can then be made from $\omega_p$ via the Drude model of conduction in metals as $n = \epsilon_0 m^* \omega_p^2 / e^2$. Inserting $\omega_p$ into this expression, assuming $m^*/m_e = 1$, results in a carrier density of $n \approx 10^{24}$ electrons / m$^3$, revealing LiOsO$_3$ to be a low carrier density metal. An estimate for $E_F$ can be made from the carrier density by assuming the weakly interacting Fermi gas relation,

$$E_F = (3\pi^2)^{2/3} \frac{\hbar^2}{2m^*} n^{2/3} \quad (5)$$

which results in a Fermi energy of only $E_F \approx 4$ meV.

With $\omega_p$ and $E_F$ determined, we can now estimate the electron-electron scattering rate $K$. Inserting $E_F$ and $\omega_p$ into Supplementary Equation 4 gives an estimated electron-electron scattering rate of $K \approx 4700$ fs$^{-1}$ eV$^{-2}$. The initial electronic temperature was determined in our three temperature model to be is $T_{e,i} \approx 270$ K - 340 K (See Supplementary Equation 7 below). At these temperatures, with the estimated electron-electron scattering rate considered, Supplementary Equation 3 suggests that excited photocarriers are expected to thermalize within $t_{\text{th}} \approx 1$ fs after initial excitation, a typical timescale for thermalization in metals. As this is significantly shorter than the duration of our pump pulse, $t_p \approx 100$ fs, excited photocarriers are expected to thermalize well before our pump pulse has left the sample, thereby satisfying the necessary initial conditions of the three-temperature model.

While the calculation presented above was performed assuming a non-interacting system, it should be noted that the true strength of electronic correlations in LiOsO$_3$ is currently an open question[8,10]. However, the calculation presented above remains valid even if electronic correlations are large, as a larger $m^*$ would only increase $K$ and therefore further decrease $t_{\text{th}}$.

## Supplementary Note 8: Details of the three temperature model

Construction of the three-temperature model begins by inputting reported values for the relevant physical properties of LiOsO$_3$. Supplementary Figures 5a-c show the temperature dependent **a**, unit cell volume, **b**, total lattice heat capacity, and **c**, resistivity as reported by Shi *et al.*[6]. Supplementary Figure 5d displays the temperature dependent skin depth at our probe wavelength of $\lambda = 1350$ nm which was calculated from the reported resistivity via the relation,

$$\delta_s = \sqrt{\frac{2\rho}{\omega\mu}} \sqrt{\sqrt{1+(\rho\omega\epsilon)^2} + \rho\omega\epsilon} \qquad (6)$$

where $\mu$ and $\epsilon$ are the relative permeability and permittivity of LiOsO$_3$ respectively. As the magnetic susceptibility of LiOsO$_3$ shows no signs of magnetic ordering[6], we assume $\mu = 1$ in our model. The relative permittivity was assumed to be $\epsilon \approx 10$, a typical value for conventional metals.

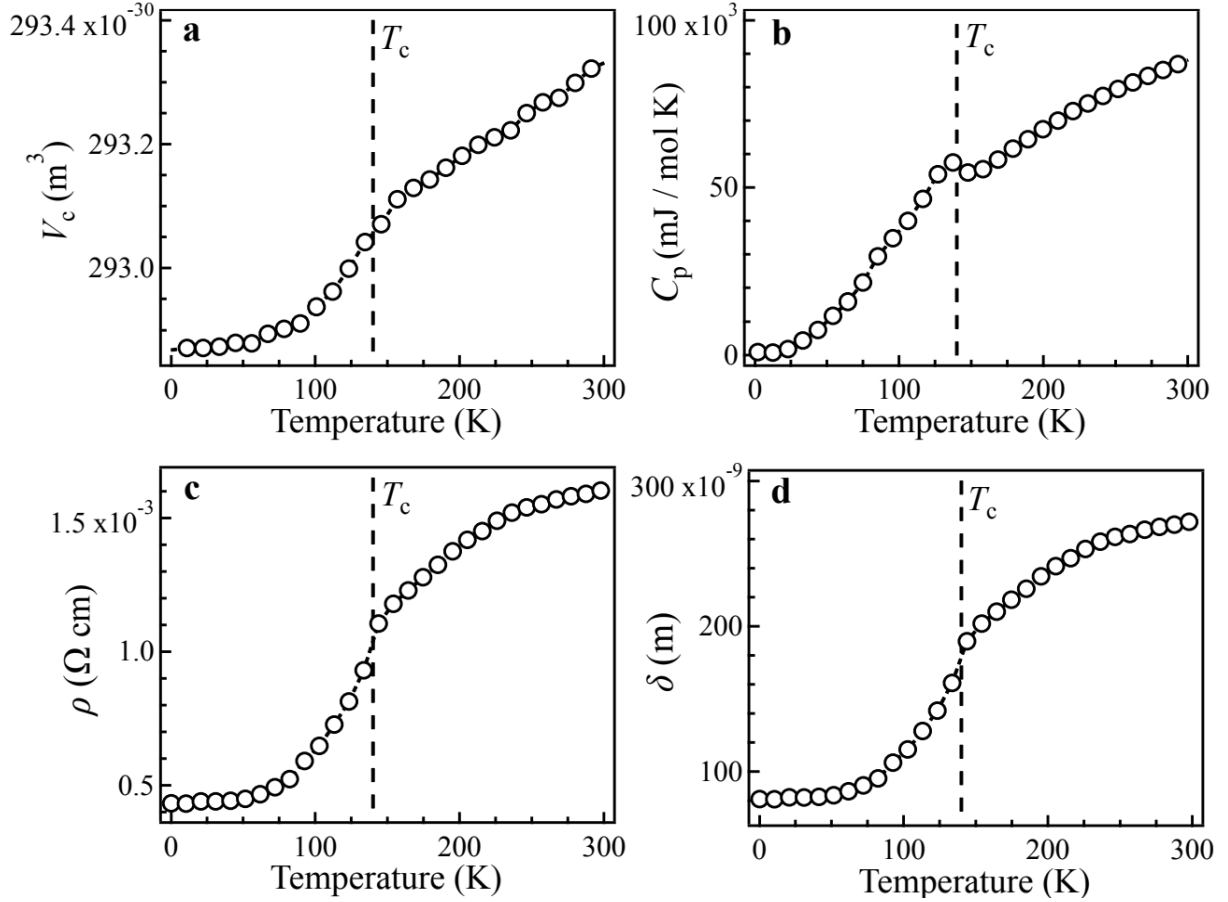

Supplementary Figure 5: **Inputs of the three temperature model** Temperature dependent **a**, unit cell volume, **b**, heat capacity, and **c**, resistivity reported by Shi. *et al.*[6]. **d**, Calculated skin depth at our probe pulse wavelength of $\lambda = 1350$ nm from the resistivity via Supplementary Equation 6.

The model assumes that the excited photocarriers strongly couple to only a subset of the phonon spectrum, which then in turn decay anharmonicly to the remainder of the phonon spectrum. The heat capacity of the excited photocarriers is assumed to be linear in temperature throughout our model and given by $C_e = \gamma T_e$ with $\gamma = 7.72$ mJ/mol K$^2$ as reported by Shi *et al.*[6]. The heat capacities of the strongly coupled phonons (SCPs) and the weakly coupled phonons (WCPs) are constructed by partitioning the total lattice heat capacity $C_p$ as $C_s = \alpha C_p$ and $C_w = (1-\alpha)C_p$ respectively, where the parameter $\alpha < 1$. In constructing our model we found that the equilibrium temperature of the lattice and photocarriers was sometimes above the temperature range of the heat capacity measurement of Shi *et al.*[6]. The total lattice heat capacity was extrapolated to higher temperatures by fitting the heat capacity outside the transition region with a third order polynomial function and extending this fit to $T = 350$ K, similar to the fits presented in Figure 3(a) of Ref. 6. Above $T = 350$ K, the lattice heat capacity was assumed to be temperature independent with a value of $C_p = 9.5 \times 10^5$ mJ / mol K, in accordance with the law of Dulong-Petit.

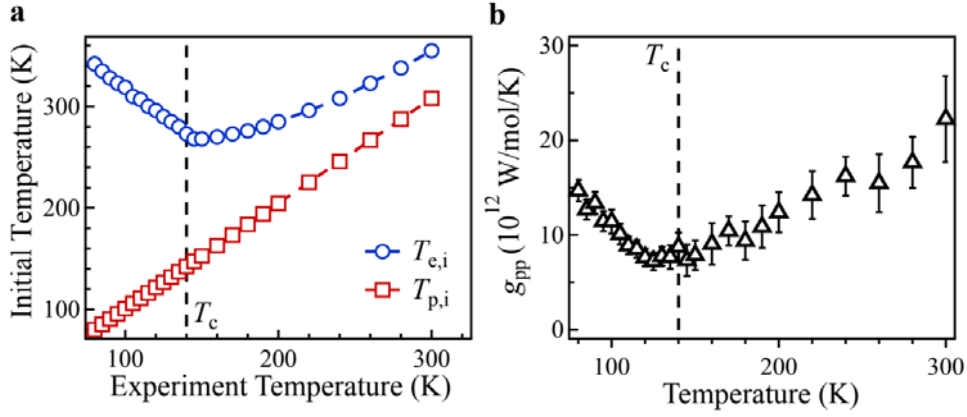

Supplementary Figure 6: **Initial temperatures and phonon-phonon coupling of the three-temperature model a**, Initial electronic (blue circles) and lattice (red squares) temperatures as a function of the initial experiment temperature in the three temperature model. **b**, The phonon-phonon coupling $g_{pp}$ extracted from the three-temperature model. Error bars derive from the $\chi^2$ of the three-temperature model fits.

The initial conditions of the model are the initial electronic and lattice temperatures immediately after excitation by the pump pulse. The initial lattice temperatures are simply the temperature of the experiment. The initial electronic temperatures are given by[11],

$$T_{e,i} = \frac{1}{\delta_s} \int_0^{\delta_s} \left[ \sqrt{T_i^2 + \frac{2(1-R)F}{\delta_s \gamma} \exp(-z/\delta_s)} \right] dz \qquad (7)$$

where $F = 0.5$ mJ/cm$^2$ is the pump pulse fluence, $R$ is the reflectivity at the pump wavelength, $T_i$ is the experiment temperature, $z$ is the depth into the sample, and integration is performed over one penetration depth $\delta_s$. We estimate that the reflectivity $R \approx 0.1$ and is very nearly temperature independent at our pump pulse wavelength of $\lambda = 800$ nm from the reflectivity experiments of Lo Vecchio et al.[8]. Supplementary Figure 6a displays the initial temperatures of the electronic (blue circles) and lattice (red squares) subsystems in our three temperature model.

Relaxation then occurs through heat exchange between the electronic and two lattice thermal baths described as captured by the equations,

$$2C_e \frac{\partial T_e}{\partial t} = -g_{ep}(T_e - T_s) + I(t,z) + \nabla \cdot [\kappa_e \nabla T_e] \qquad (8)$$

$$C_s \frac{\partial T_s}{\partial t} = g_{ep}(T_e - T_s) - g_{pp}(T_s - T_w) \qquad (9)$$

$$C_w \frac{\partial T_w}{\partial t} = g_{pp}(T_s - T_w) \qquad (10)$$

where $T_e$, $T_s$, and $T_w$ are the time dependent electronic and lattice temperatures respectively and $g_{ep}$ and $g_{pp}$ are the electron-phonon and phonon-phonon couplings. $I(t,z)$ is the laser source term given by,

$$I(t,z) = \frac{2ln(2)(1-R)F}{\pi t_p \delta_s} \exp\left[-4ln(2)\left(\frac{t}{t_p}\right)^2 - \frac{z}{\delta_s}\right] \qquad (11)$$

where $t_p = 100$ fs is the pulse duration. $R$ in this case now refers to the reflectivity at our probe pulse wavelength of $\lambda = 1350$nm, which we estimate to be $R = 0.35$ from Lo Vecchio et al.[8]. The last term in Supplementary Equation 8 describes heat diffusion out of the probed region of the sample. An estimate for the thermal conductivity $\kappa_e$ was obtained by assuming LiOsO$_3$ obeys the Wiedemann-Franz law which relates $\kappa_e$ to the resistivity as $\kappa_e = LT_e/\rho$, where $L = 2.44 \times 10^{-8}$ W$\Omega$K$^{-2}$ is the Lorentz number. Using this relation we estimate the thermal conductivity of LiOsO$_3$ to be $\kappa_e = 0.4$ W/m/K at $T_e = 100$K.

Solving the model under the experimentally dictated initial conditions gives the transient electronic and lattice temperatures, which are displayed as a function of initial experiment temperature in Supplementary Figure 7. Model reflectivity transients were then constructed by assuming the conventional[11,12] linear combination of the electronic and lattice temperatures as,

$$\frac{\Delta R}{R} = aT_e + b[\alpha T_s + (1-\alpha)T_w] \quad (12)$$

where $a$ and $b$ are determined by matching the initial and final values of the experimental reflectivity transients. These model reflectivity transients were then convolved with a normalized Gaussian of the form,

$$g(t) = \frac{1}{t_p\sqrt{2\pi}} \exp\left[-\left(\frac{t}{\sqrt{2}t_p}\right)^2\right] \quad (13)$$

where again $t_p = 100$ fs is our pulse duration, to simulate the effects of our pump and probe pulses. Finally, the model reflectivity transients were then fit to the experimental data via a least squares regression algorithm with $g_{ep}$, $g_{pp}$ and $\alpha$ as fully relaxed fitting coefficients.

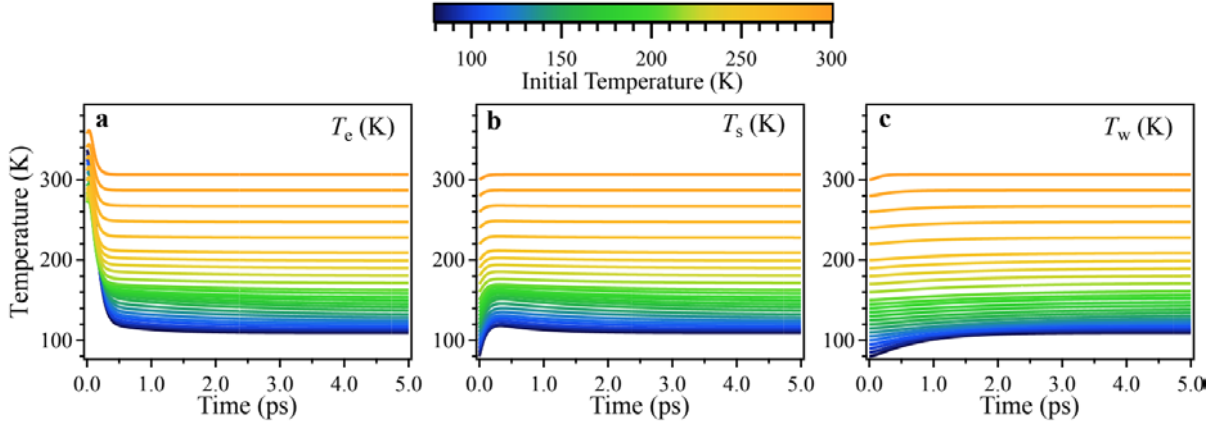

Supplementary Figure 7: **Transient temperatures of the three-temperature model** Solutions of the three-temperature model at various initial experiment temperatures as indicated by the color scale at the top. Panels **a-c** display the transient **a**, electronic $T_e$, **b**, strongly coupled phonon $T_s$, and **c**, weakly coupled phonon $T_w$ temperatures respectively.

Supplementary Figure 6b displays the phonon-phonon coupling constant $g_{pp}$ extracted from the three-temperature model. The phonon-phonon coupling can be seen to decrease as the polar transition is approached, suggesting that the lattice becomes overall more harmonic as the temperature is reduced. However, $g_{pp}$ increases in the polar phase. It is unclear where this increase originates but it is possible that such an increase could stem from relaxed phonon coupling selection rules in the polar phase due to broken inversion symmetry, which would permit additional coupling channels that were symmetry forbidden above the transition.

## Supplementary Note 9: Comparison of the electron-phonon coupling function to the $E_g$ phonon linewidths

First principles electron-phonon coupling theory predicts that the linewidths of coupled phonons scale with the electron-phonon coupling function as $g_{ep} \propto (\hbar \Gamma_{ph})^{1/2}$ in the zero momentum limit[13]. Such a comparison was vital in determining which phonon modes constitute the strongly coupled phonons in LiOsO$_3$. Supplementary Figure 8 shows a comparison between our extracted temperature dependent electron-phonon coupling and the scaled square roots of the linewidths of the **a**, $^1E_g$, **b**, $^2E_g$, and **c**, average $E_g$ phonon modes as labeled by their representations in the centrosymmetric $R\bar{3}c$ space group and reported by Jin et al.[14]. One can see that the temperature dependence of both quantities display strong similarities. Their correlation is better observed by plotting the normalized electron-phonon coupling versus the normalized square roots of the $E_g$ phonon linewidths, which are shown in Supplementary Figures 8**d-f**. The black dashed lines represent the limit of perfect correlation $g_{ep} \propto \Gamma_{ph}^{1/2}$. These plots demonstrate the strong correlation between these two quantities.

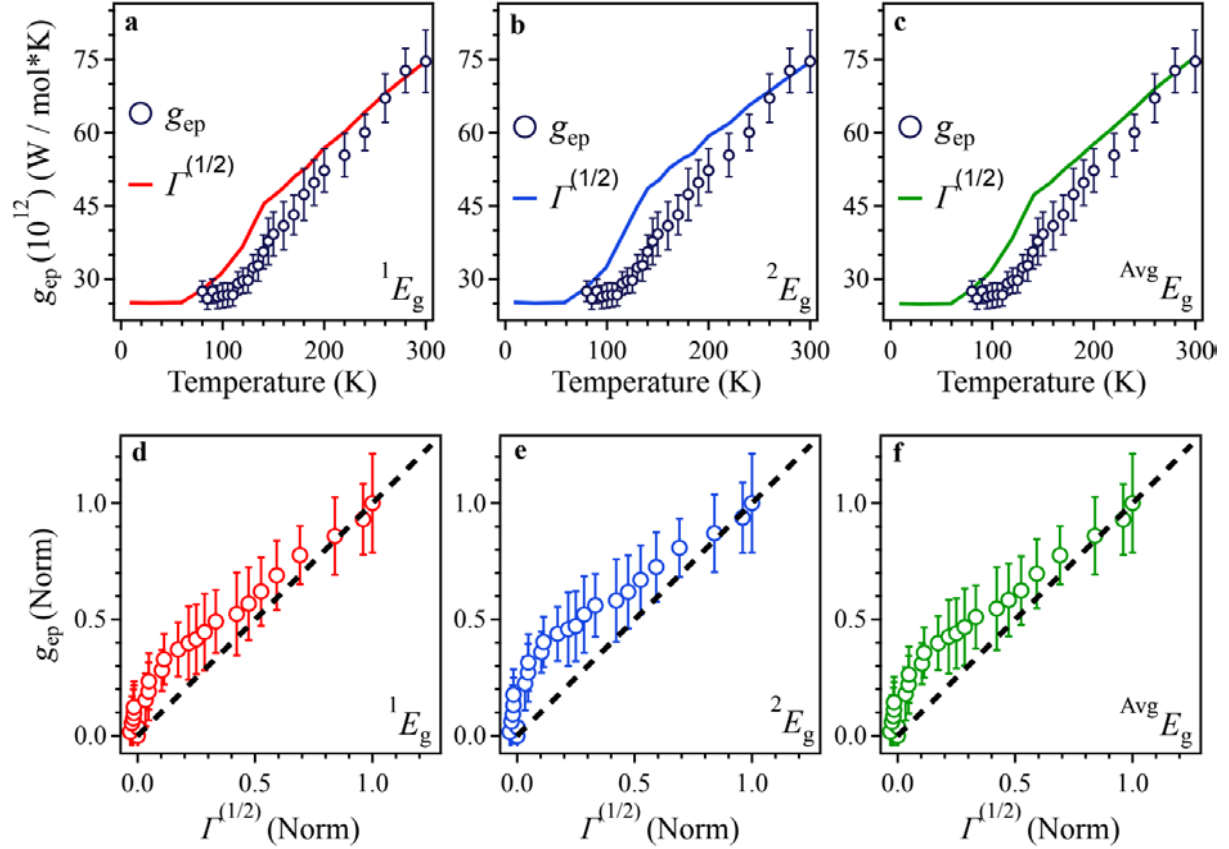

Supplementary Figure 8: **Identification of the strongly coupled phonons** The temperature dependent electron-phonon coupling function $g_{ep}$ extracted from our three-temperature model plotted with the scaled square roots of the **a**, $^1E_g$, **b**, $^2E_g$, and **c**, average $E_g$ phonon linewidths reported via Raman spectroscopy[14]. **d-f**, Plots of the normalized $g_{ep}$ versus the normalized square roots of the $E_g$ phonon linewidths showing a strong correlation between these two quantities. Black dashed lines represent the limit of perfect correlation $g_{ep} \propto \Gamma_{ph}^{1/2}$. Error bars derive from the $\chi^2$ of the three-temperature model fits.

## Supplementary Note 10: Details of the displacive soft mode model of the weakly coupled heat capacity

From the three-temperature model analysis, we concluded that the itinerant electrons of $LiOsO_3$ are decoupled from a significant portion of the total lattice heat capacity. We argued that at temperatures $T < 250$ K, at which point the other optical phonon modes are expected to be largely frozen out, the decoupled heat capacity can be mostly attributed to that of the $A_{2u}$ polar mode. That is,

$$1 - \alpha \approx C_{A_{2u}}/C_p \quad (14)$$

where $C_{A_{2u}}$ is the heat capacity of the $A_{2u}$ polar mode and $C_p$ is the temperature dependent total lattice heat capacity of $LiOsO_3$ as reported by Shi et al.[6].

To model $1 - \alpha$, the polar mode was treated as an Einstein phonon whose temperature dependent heat capacity is given by

$$C_{A_{2u}} = A\left(\frac{\hbar\omega}{k_B T}\right) \frac{e^{\hbar\omega/k_B T}}{[e^{\hbar\omega/k_B T} - 1]^2} \quad (15)$$

where $A$ is a proportionality constant and $\hbar\omega$ is the energy of the polar mode. Supplementary Equation 15 was inserted into Supplementary Equation 14 and fit to $1 - \alpha$ with the phonon frequency as the only free parameter. The proportionality constant $A$ was chosen such that the polar mode energy softened to zero at the transition as expected for a displacive mode. Fitting the data in this fashion allows one to extract the temperature dependent heat capacity and frequency of the polar mode which was shown in the inset of Figure 4d of the main text.

## Supplementary Note 11: Phonon dispersions of the polar R3c phase

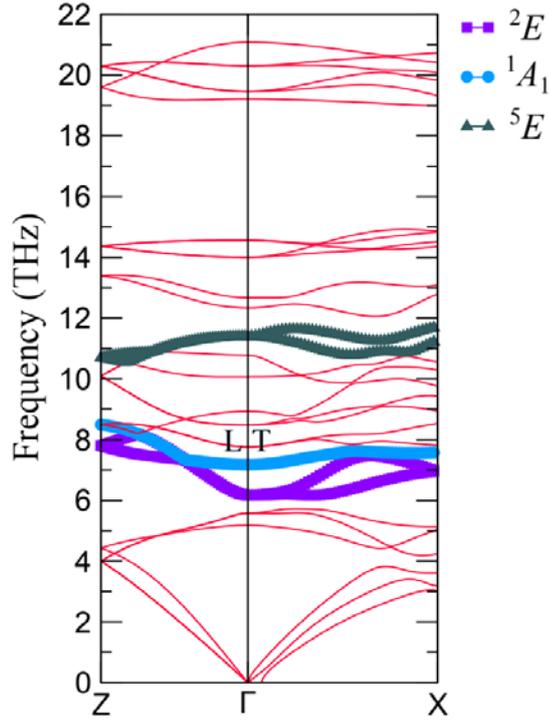

Supplementary Figure 9: **Phonon dispersions of the R3c phase of $LiOsO_3$** The polar $A_{2u}$ mode and the doubly degenerate $^1E_g$ and $^2E_g$ modes, which transform as $^1A_1$, $^2E$, and $^5E$ respectively in the polar phase in accordance with the notation of[14], are highlighted. Only the polar mode can be defined as longitudinal (L) or transverse (T) in the Brillouin zone, as the other modes exhibit both longitudinal and transverse components throughout the Brillouin zone.

## Supplementary References